\documentclass[aps,twocolumn,showpacs,groupedaddress]{revtex4}

\usepackage{graphicx}
\usepackage{dcolumn}
\usepackage{bm}
\usepackage{amssymb}

\usepackage[usenames]{color}

\begin{document}

\newcommand{\dzero}     {D0}
\newcommand{\met}       {\mbox{$\not\!\!E_T$}}
\newcommand{\deta}      {\mbox{$\eta^{\rm det}$}}
\newcommand{\meta}      {\mbox{$\left|\eta\right|$}}
\newcommand{\mdeta}     {\mbox{$\left|\eta^{\rm det}\right|$}}
\newcommand{\rar}       {\rightarrow}
\newcommand{\rargap}    {\mbox{ $\rightarrow$ }}
\newcommand{\tbbar}     {\mbox{$tb$}}
\newcommand{\tqbbar}    {\mbox{$tqb$}}
\newcommand{\ttbar}     {\mbox{$t\bar{t}$}}
\newcommand{\bbbar}     {\mbox{$b\bar{b}$}}
\newcommand{\ccbar}     {\mbox{$c\bar{c}$}}
\newcommand{\qqbar}     {\mbox{$q\bar{q}$}}
\newcommand{\ppbar}     {\mbox{$p\bar{p}$}}
\newcommand{\comphep}   {\sc{c}\rm{omp}\sc{hep}}
\newcommand{\herwig}    {\sc{herwig}}
\newcommand{\pythia}    {\sc{pythia}}
\newcommand{\alpgen}    {\sc{alpgen}}
\newcommand{\singletop} {\rm{SingleTop}}
\newcommand{\reco}      {\sc{reco}}
\newcommand{\Mchiggs}   {\mbox{$M({\rm jet1,jet2},W)$}}
\newcommand{\coss}	{\mbox{\rm{cos}$\theta^{\star}$}}
\newcommand{\ljets} {$\ell+$jets}

\lefthyphenmin=6
\righthyphenmin=6

\hspace{5.2in}\mbox{Fermilab-Pub-09/372-E}



\title{Measurement of the t-channel single top quark production cross section} 
%
\author{V.M.~Abazov$^{37}$}
\author{B.~Abbott$^{75}$}
\author{M.~Abolins$^{65}$}
\author{B.S.~Acharya$^{30}$}
\author{M.~Adams$^{51}$}
\author{T.~Adams$^{49}$}
\author{E.~Aguilo$^{6}$}
\author{M.~Ahsan$^{59}$}
\author{G.D.~Alexeev$^{37}$}
\author{G.~Alkhazov$^{41}$}
% ...

\affiliation{\vspace{0.1 in}(The D\O\ Collaboration)\vspace{0.1 in}}
% ...
\affiliation{$^{6}$University of Alberta, Edmonton, Alberta, Canada;
                Simon Fraser University, Burnaby, British Columbia, Canada;
                York University, Toronto, Ontario, Canada and
                McGill University, Montreal, Quebec, Canada}
% ...
\affiliation{$^{30}$Tata Institute of Fundamental Research, Mumbai, India}
% ...
\affiliation{$^{37}$Joint Institute for Nuclear Research, Dubna, Russia}
% ...
\affiliation{$^{41}$Petersburg Nuclear Physics Institute,
                St. Petersburg, Russia}
% ...
\affiliation{$^{49}$Florida State University, Tallahassee, Florida 32306, USA}
% ...
\affiliation{$^{51}$University of Illinois at Chicago,
                Chicago, Illinois 60607, USA}
% ...
\affiliation{$^{59}$Kansas State University, Manhattan, Kansas 66506, USA}
% ...
\affiliation{$^{65}$Michigan State University,
                East Lansing, Michigan 48824, USA}
% ...
\affiliation{$^{75}$University of Oklahoma, Norman, Oklahoma 73019, USA}
% ...

\date{July 24, 2009}

\begin{abstract}
The D0 collaboration reports direct evidence for electroweak production of
single top quarks through the $t$-channel exchange of a virtual $W$~boson. This is the first
analysis to isolate an individual single top quark production channel. We select events 
containing an isolated electron or muon, missing transverse energy, and two, three or four jets 
from 2.3~fb$^{-1}$ of {\ppbar} collisions at the Fermilab Tevatron Collider. 
One or two of the jets are identified as containing a $b$~hadron. 
We combine three multivariate techniques optimized for the $t$-channel process to measure the 
$t$-~and $s$-channel cross sections simultaneously. We measure cross sections of
$3.14^{+0.94}_{-0.80}$~pb for the $t$-channel and $1.05 \pm 0.81$~pb for the $s$-channel. 
The measured $t$-channel result is found to have a significance of 4.8 standard deviations and 
is consistent with the standard model prediction.
\end{abstract}



\maketitle



\begin{figure}[h!t]
\centering
\includegraphics[width=0.235\textwidth]{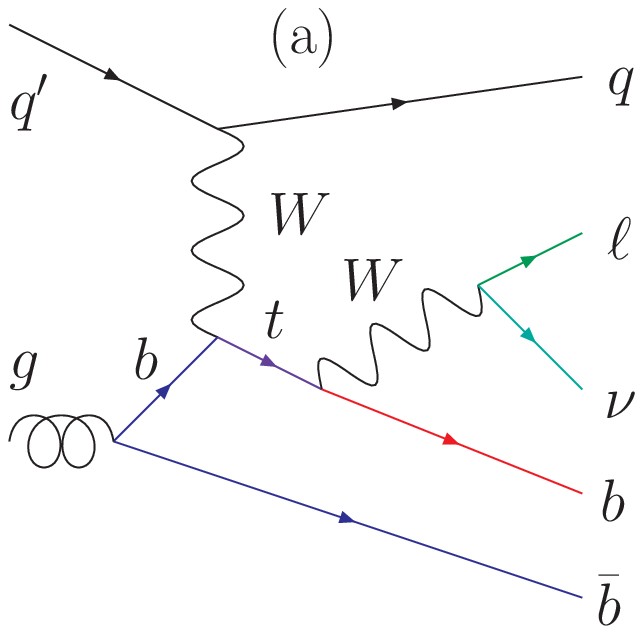}
\hspace{-0.2in}
\includegraphics[width=0.26\textwidth]{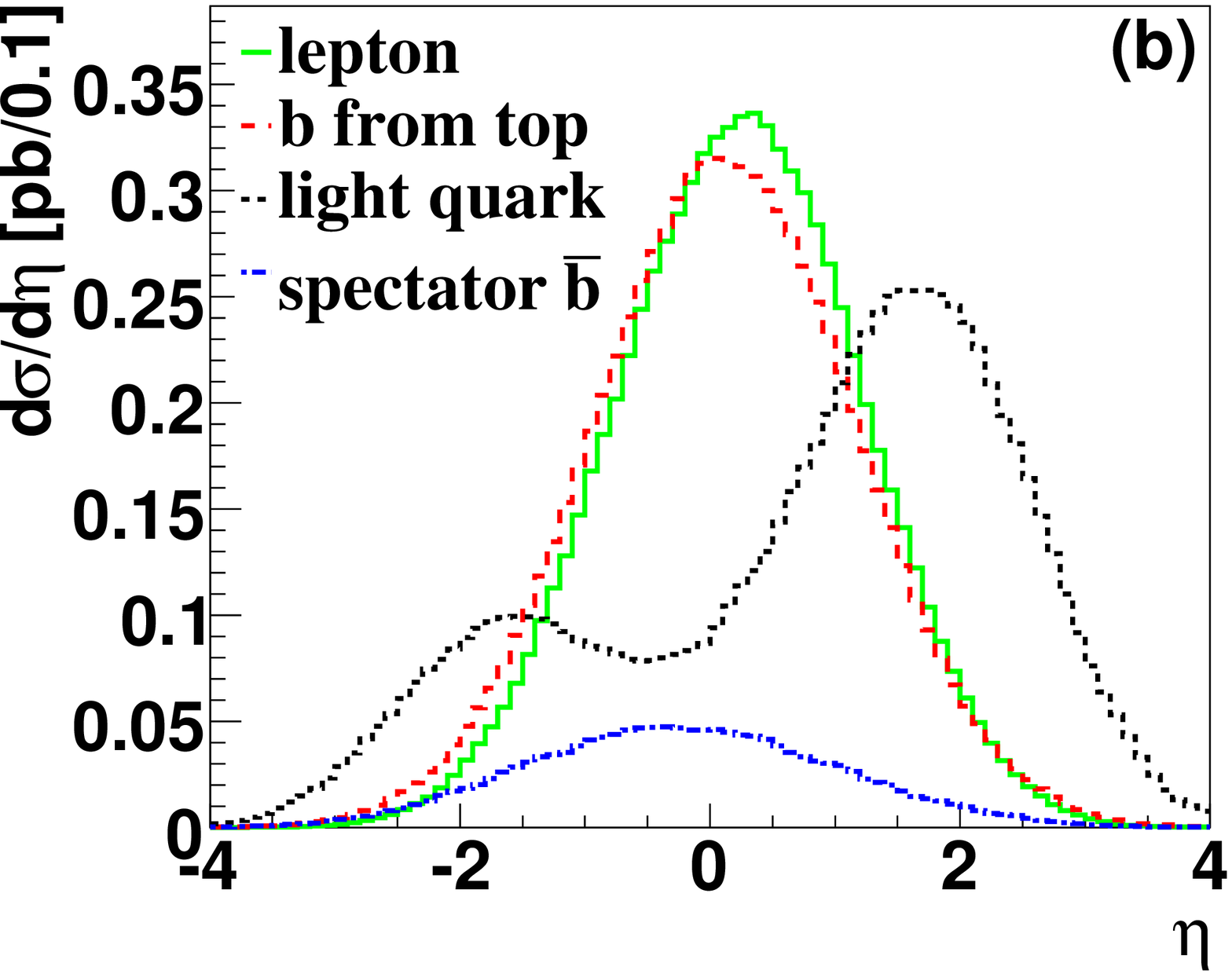}
\vspace{-0.3in}
\caption{Representative Feynman diagram for $t$-channel single top quark
production and decay (a) and parton-level pseudorapidity distribution of the final
state objects in top production (excluding antitop), requiring each object to have transverse 
momentum $>15$~GeV (b).}
\label{fig:feynman}
\end{figure}

\vspace{-0.1in}
The D0 and CDF collaborations at the Fermilab Tevatron {\ppbar} Collider
have recently observed electroweak production of
single top quarks~\cite{singletop-obs-d0,Aaltonen:2009jj}, measuring
the total single top quark production cross section as well as the 
Cabibbo-Kobayashi-Maskawa (CKM) matrix element~\cite{Cabibbo:1963yz} 
$|V_{tb}|$ directly. In the standard model (SM), the two main production modes at the Tevatron 
resulting in a single top (or antitop) quark final state are the $t$-channel exchange 
of a $W$~boson shown in Fig.~\ref{fig:feynman}a and
the $s$-channel production via the decay of a virtual $W$~boson. The two
observation analyses measured only the combined single top quark cross section, assuming the
SM ratio of the two production modes. This ratio is modified in several
new physics scenarios, for example in models with additional quark generations,
new heavy bosons~\cite{Tait:2000sh,Abazov:2008vj,Abazov:2006aj,Aaltonen:2009qu}, 
flavor-changing neutral currents~\cite{Hosch_Whishant,d0-fcnc,Aaltonen:2008qr}, or
anomalous top quark 
couplings~\cite{Chen:2005vr,dudko-boos,singletop-wtb-heinson,d0-singletop-wtb,Abazov:2009ky}.
In this Letter we remove this assumption and use the $t$-channel characteristics 
to measure the $t$-channel and $s$-channel cross sections simultaneously, thus providing
a $t$-channel measurement independent of the $s$-channel cross section model. The main
characteristic of the $t$-channel which separates it both from the $s$-channel and the
backgrounds is the pseudorapidity distribution of the 
light quark jet, shown in Fig.~\ref{fig:feynman}b.
The predicted cross section
for $t$-channel ($s$-channel) production is $2.34 \pm 0.13$~pb 
($1.12 \pm 0.04$~pb) for a top quark mass $m_t=170$~GeV~\cite{singletop-xsec-kidonakis}. 
\vspace{-0.05in}
\vspace{-0.1in}

This analysis extends the D0 single top evidence~\cite{d0-prl-2007,d0-prd-2008} 
and observation analyses~\cite{singletop-obs-d0}, utilizing the same dataset, event
selection, and signal/background modeling as the observation analysis, but training 
multivariate filters specifically to extract $t$-channel single top quark events.
We use 2.3~fb$^{-1}$ of data collected by the 
D0 experiment~\cite{NIM} at the Fermilab Tevatron $p\bar{p}$ Collider between 2002 and 
2007 (Run~II). 
The measurement selects final states containing one high transverse
momentum ($p_T$) isolated lepton (electron or muon), large missing
transverse energy ({\met}), a $b$~quark jet from the decay of the top
quark ($t{\rar}Wb{\rar}\ell \nu b$), a light quark jet produced in association with the top 
quark, and a spectator $b$~quark jet from gluon splitting in the initial state. 
We allow for one of these jets not to be 
identified as well as for the presence of an additional jet from gluon radiation.
The backgrounds are $W$~bosons
produced in association with jets, {\ttbar} pairs, and multijet
production, where a jet is misreconstructed as an electron or a
heavy-flavor quark decays to a muon that satisfies isolation
criteria. $Z$+jets and diboson processes form minor additional background components. 
We treat $s$-channel single top quark production
as a background during the multivariate training but measure its cross section 
simultaneously with the $t$-channel measurement as explained below.

We look for $t$-channel and $s$-channel single top quark production in events with two to 
four jets with $p_T>15$~GeV
and pseudorapidity $|\eta|<3.4$, with the leading jet additionally satisfying $p_T>25$~GeV. 
 We require $20 < {\met} < 200$~GeV for events with two jets and $25 < {\met} < 200$~GeV for
events with three or four jets. Events must contain only one isolated
electron with $p_T>15$~GeV and $|\eta|<1.1$ ($p_T>20$~GeV for three-
or four-jet events), or one isolated muon with $p_T>15$~GeV and
$|\eta|<2.0$. The background from jets misidentified as leptons is kept to
approximately~5\% by requiring the total transverse energy of all final state objects
$H_T(\ell,{\met},{\rm jets})$ to be greater than 110 to 160~GeV, depending on
the analysis channel, and by demanding that the {\met} is not along the direction of the lepton 
or the leading jet in the transverse plane. To enhance the signal fraction, one or two of the 
jets are required to originate from $b$~hadrons, as implemented
through a neural network (NN) $b$-jet tagging
algorithm~\cite{btagging-scanlon}. 
We divide the dataset into 24 independent analysis channels 
(separated by data taking period, lepton type,
$b$-tag and jet multiplicity) and combine the results to maximize the signal sensitivity.
Details on the selection criteria and background 
modeling are given in Ref.~\cite{d0-prd-2008}.

We generate $t$-channel and $s$-channel single top events with the 
{\sc SingleTop} Monte Carlo (MC) generator~\cite{singletop-mcgen}.
The kinematics of the generated events closely match those predicted by 
next-to-leading-order (NLO) calculations~\cite{singletop-xsec-sullivan},
particularly those including NLO corrections to the 
$t$-channel $2\rar 3$ process shown in Fig.~\ref{fig:feynman}a~\cite{Campbell:2009ss}. 
The {\alpgen} leading-order MC event 
generator~\cite{alpgen}, interfaced to {\sc pythia}~\cite{pythia},
is used to model {\ttbar}, $W$+jets, and $Z$+jets 
background events. We use the CTEQ6L1 parton distribution functions (CTEQ6M for
single top)~\cite{cteq} and
set the top quark mass to 170~GeV. We use {\sc geant}~\cite{geant} to simulate
the response of the D0 detector to the MC events. The {\ttbar} background is
normalized to the predicted cross section~\cite{ttbar-xsec}. 
The $Z$+jets contributions are normalized to NLO cross sections~\cite{mcfm}. The
$W$+jets background normalization, jet flavor composition, and jet angular
distributions are obtained from data samples.
We model the background from multijet production where a jet is misidentified as an isolated 
electron or muon using events from data containing lepton candidates which pass all of the 
lepton identification requirements except one, but otherwise resemble the signal events.
We use {\pythia} to model diboson production.

%
We select 4519 lepton+jets events with at least one $b$-tagged jet, which
are expected to contain $130\pm 17$ $t$-channel ($93\pm 14$ $s$-channel) single top events with 
an acceptance of $(2.5 \pm 0.3)\%$ ($(3.7 \pm 0.5)\%$). The expected sample composition is shown 
in Table~\ref{tab:eventyields}.
\begin{table}[!h!btp]
\centering
\vspace{-0.15in}
\caption{Number of expected and observed events in
2.3~fb$^{-1}$ for $e$ and $\mu$, and one and two $b$-tagged analysis
channels combined, with uncertainties including both statistical and systematic components. 
The $t$-channel and $s$-channel contributions are normalized to their SM expectation.}
\label{tab:eventyields}
\begin{tabular*}{1.0\columnwidth}%
      {l@{\extracolsep{\fill}}r@{\extracolsep{0pt}$\:\pm\:$}l@{}%
        @{\extracolsep{\fill}}r@{\extracolsep{0pt}$\,\pm\,$}l@{}%
        @{\extracolsep{\fill}}r@{\extracolsep{0pt}$\,\pm\,$}l@{}}
~~~~Source & \multicolumn{2}{c}{2 jets} & \multicolumn{2}{c}{3 jets}
                                        & \multicolumn{2}{c}{4 jets}
\vspace{0.03in} \\
\hline
$t$-channel               &   77  &  10 &    39 &   6 &   14 &  3 \\
$s$-channel               &   62  &   9 &    24 &   4 &    7 &  2 \\
$W$+jets                  & 1829  & 161 &   637 &  61 &  180 & 18 \\
$Z$+jets and dibosons     &  229  &  38 &    85 &  17 &   26 &  7 \\
${\ttbar}{\rargap}\ell\ell$, $\ell$+jets
                          &  222  &  35 &   436 &  66 &  484 & 71 \\
Multijets                 &  196  &  50 &    73 &  17 &  30 &  6 \\ 
Total prediction~~        & 2615  & 192 &  1294 & 107 &  742 & 80 \\
Data                      & \multicolumn{2}{c}{2579}
                          & \multicolumn{2}{c}{1216}
			  & \multicolumn{2}{c}{724}      
\end{tabular*}
\vspace{-0.1in}
\end{table}

Systematic uncertainties in the signal and background models are 
described in detail in Ref.~\cite{d0-prd-2008}. The main uncertainties
are due to the jet energy scale (JES) corrections and the
tag-rate functions (TRF), with smaller contributions from
MC statistics, correction for jet-flavor composition in $W$+jets events, 
and from the $W$+jets, multijets, and
{\ttbar} normalizations. The total uncertainty on the background is
(8--16)\% depending on the analysis channel. 
Uncertainties on JES, TRFs and the modeling of $W$+jets kinematics
affect not only the normalization but also the shape of the
discriminant distributions. Since the $W$+jets background normalization
and kinematics are constrained by data, theory prediction uncertainties
for $W$+jets are negligible (including those on the parton distribution
functions and factorization and normalization scale).

\begin{figure}[!h!tbp]
\centering
\vspace{-0.1in}
\hspace{-0.1in}
\includegraphics[width=0.245\textwidth]{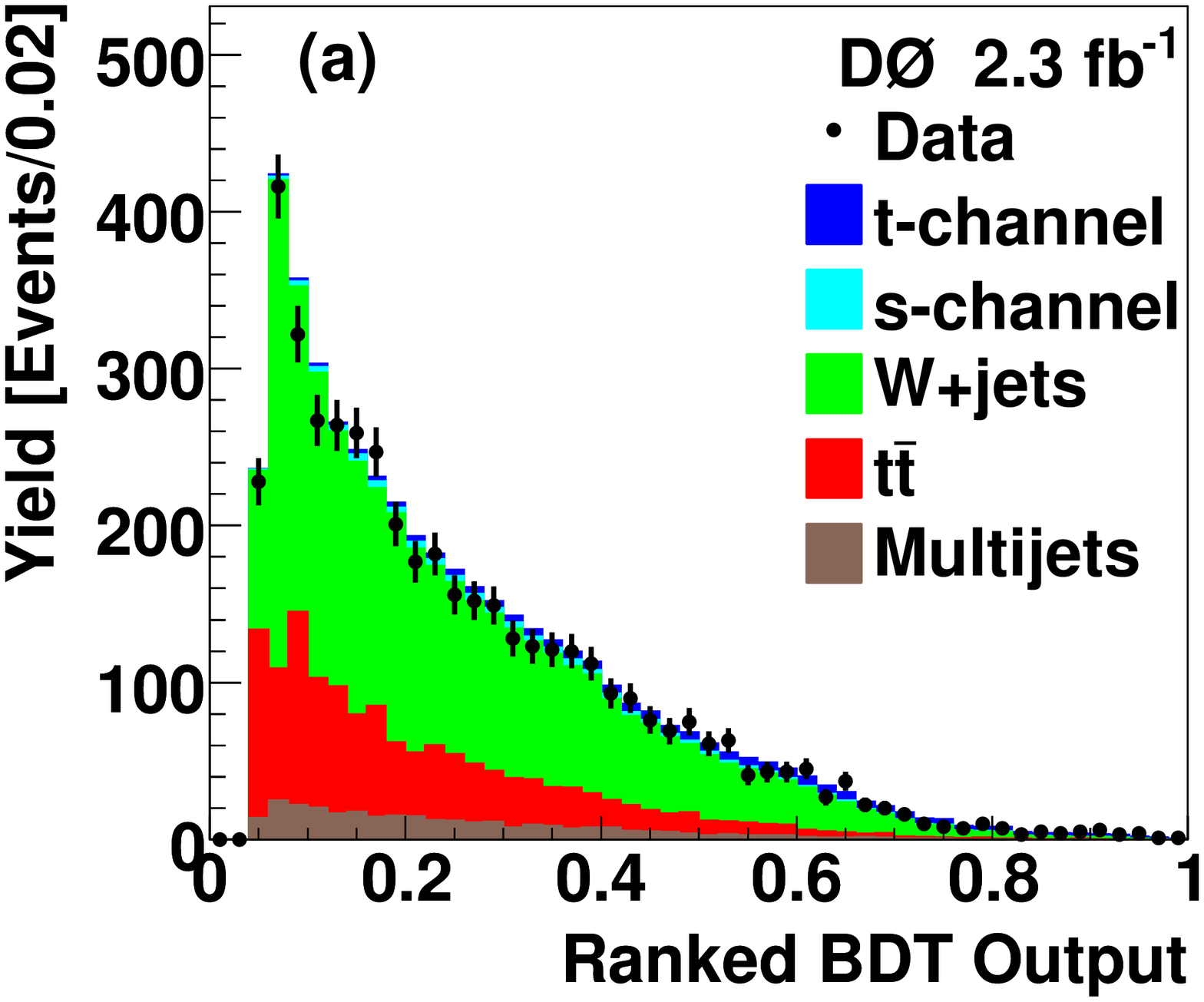}
\hspace{-0.1in}
\includegraphics[width=0.245\textwidth]{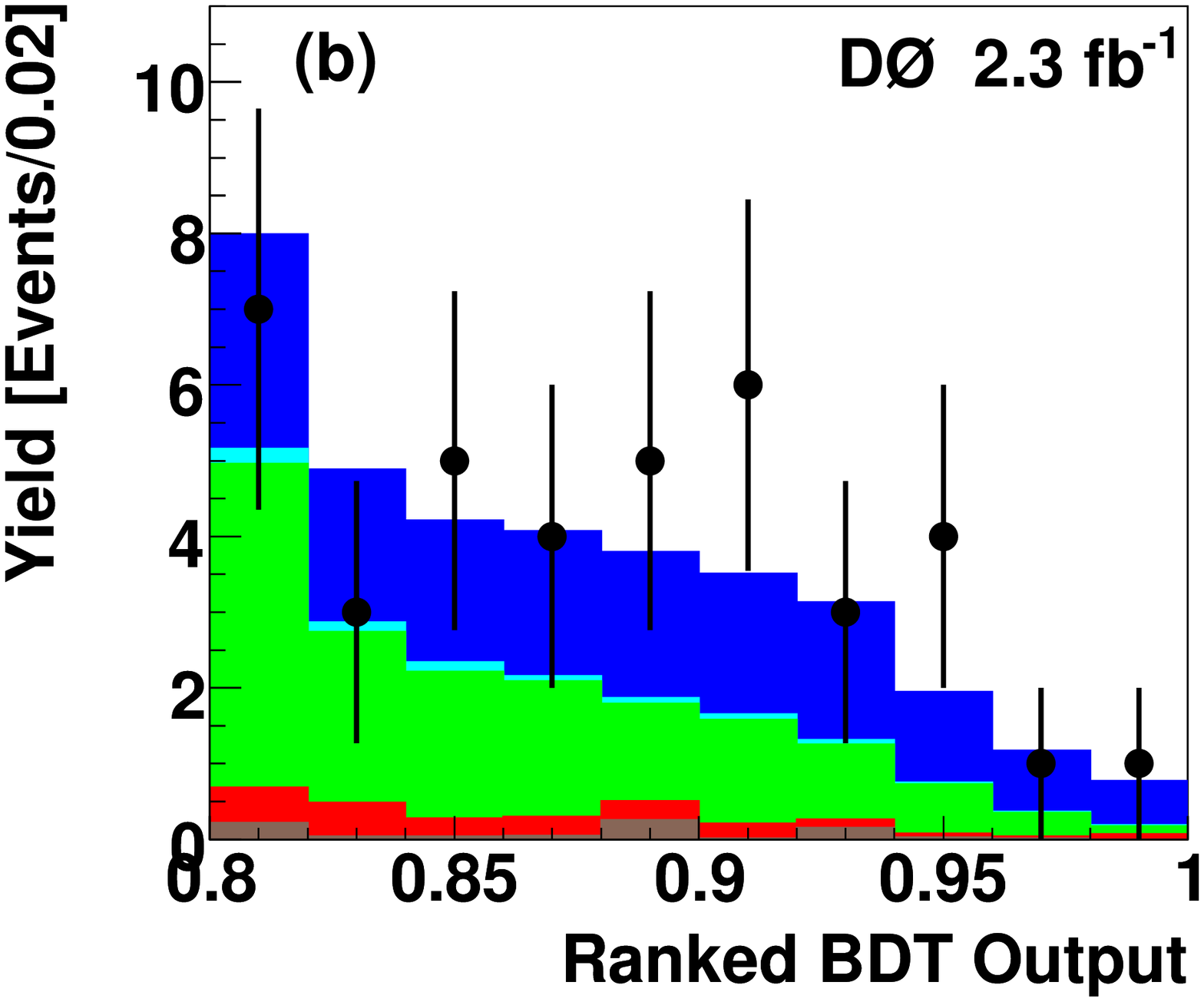}
\\
\vspace{-0.03in}
\hspace{-0.1in}
\includegraphics[width=0.245\textwidth]{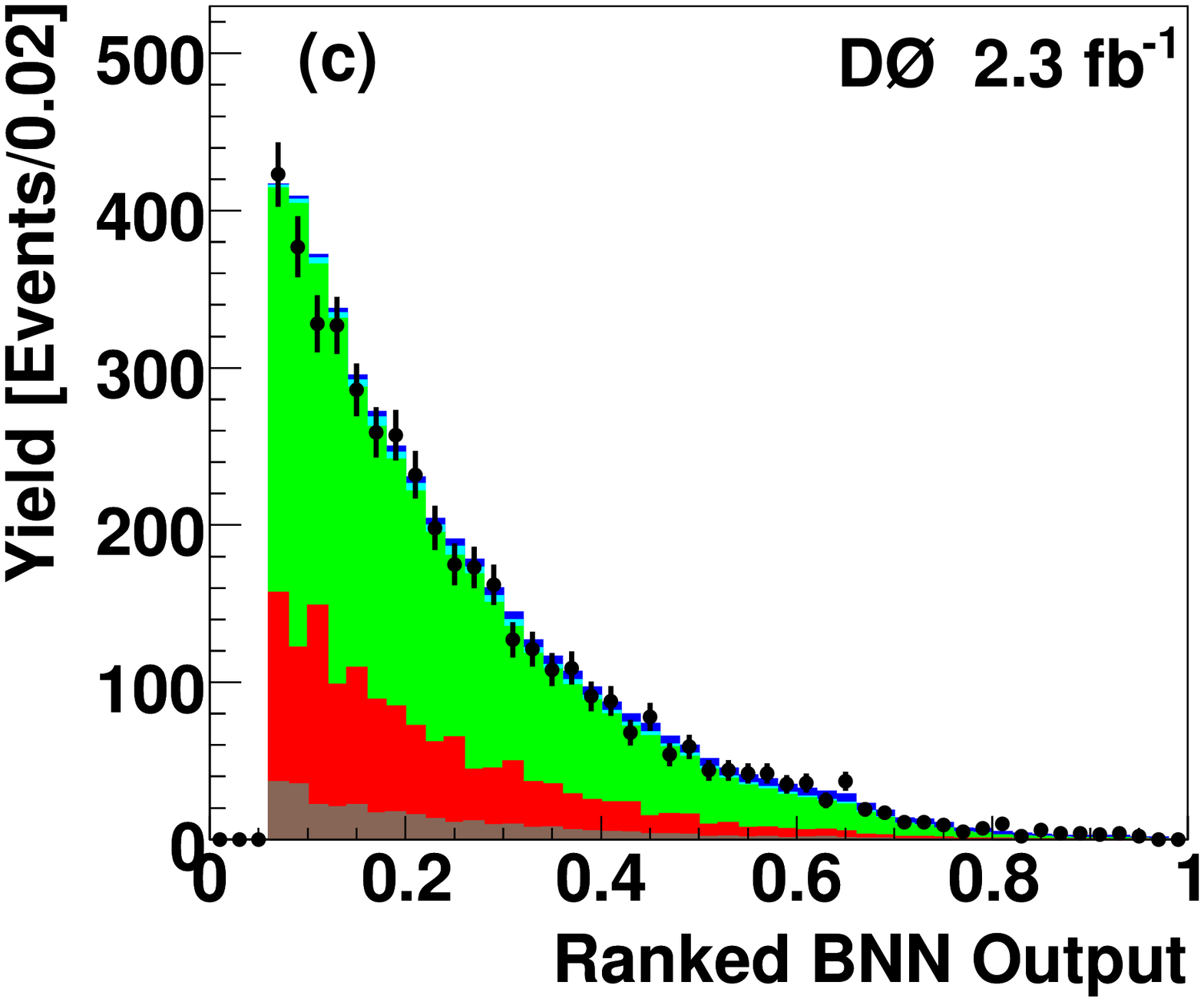}
\hspace{-0.1in}
\includegraphics[width=0.245\textwidth]{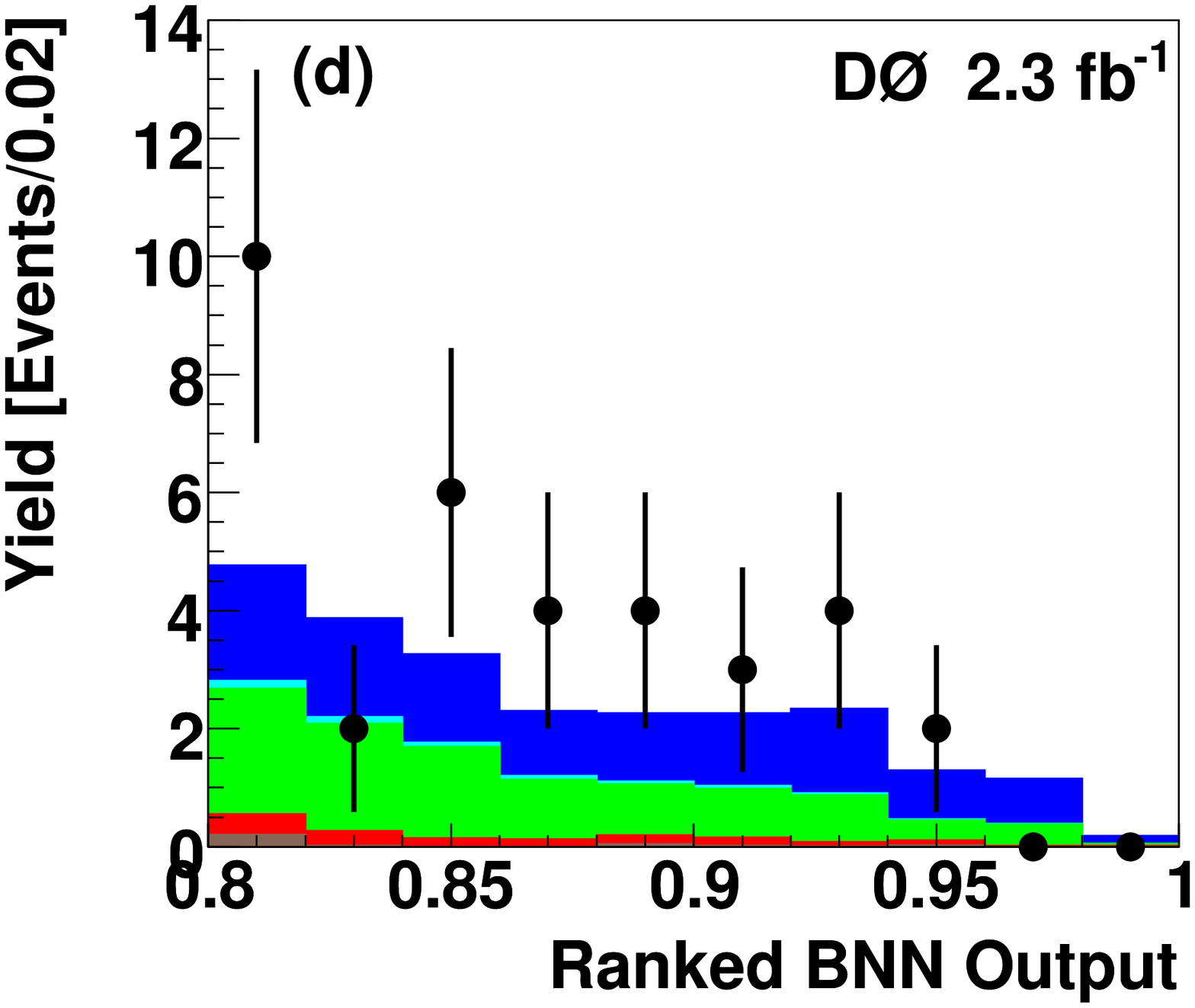}
\\
\vspace{-0.03in}
\hspace{-0.1in}
\includegraphics[width=0.245\textwidth]{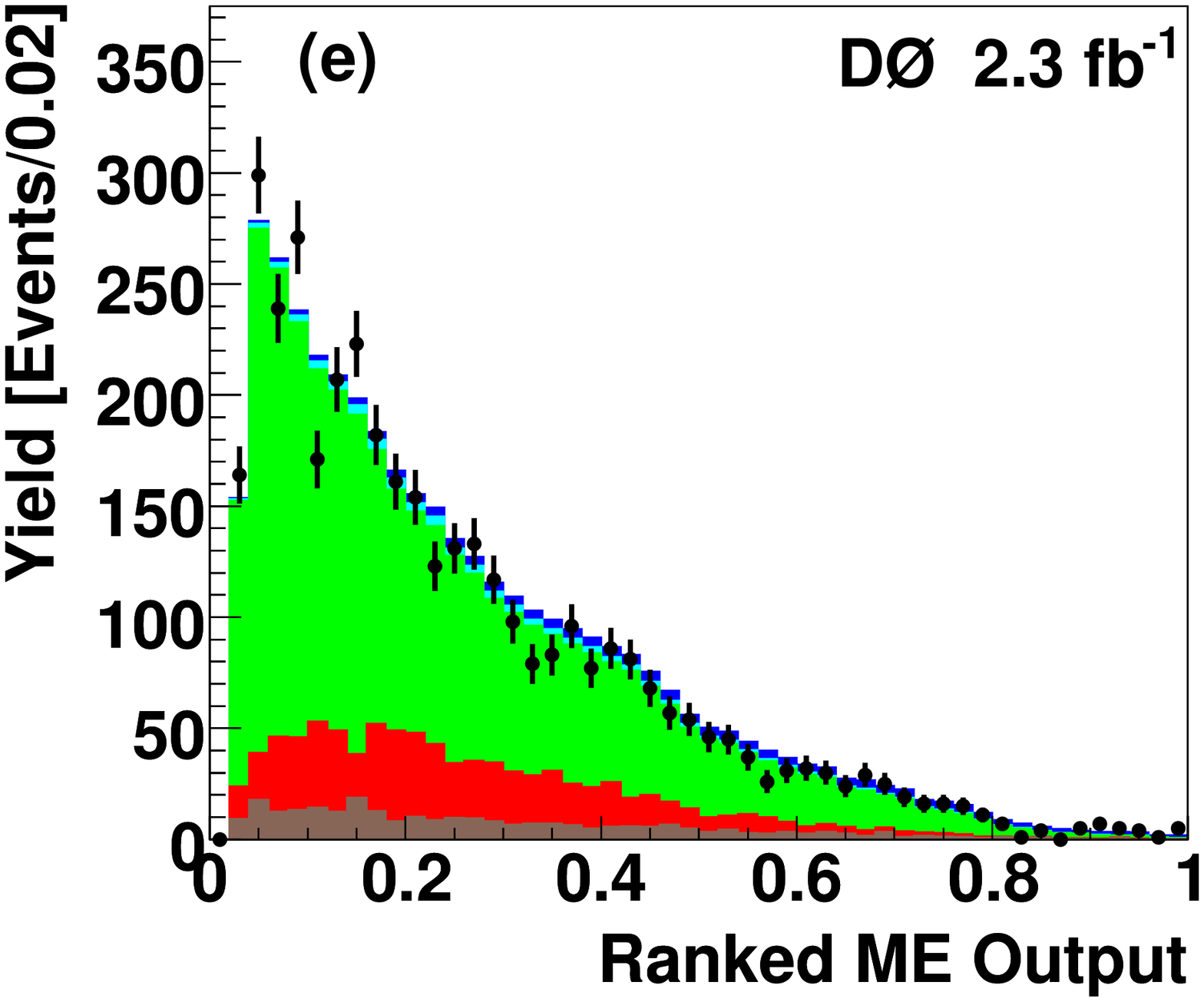}
\hspace{-0.1in}
\includegraphics[width=0.245\textwidth]{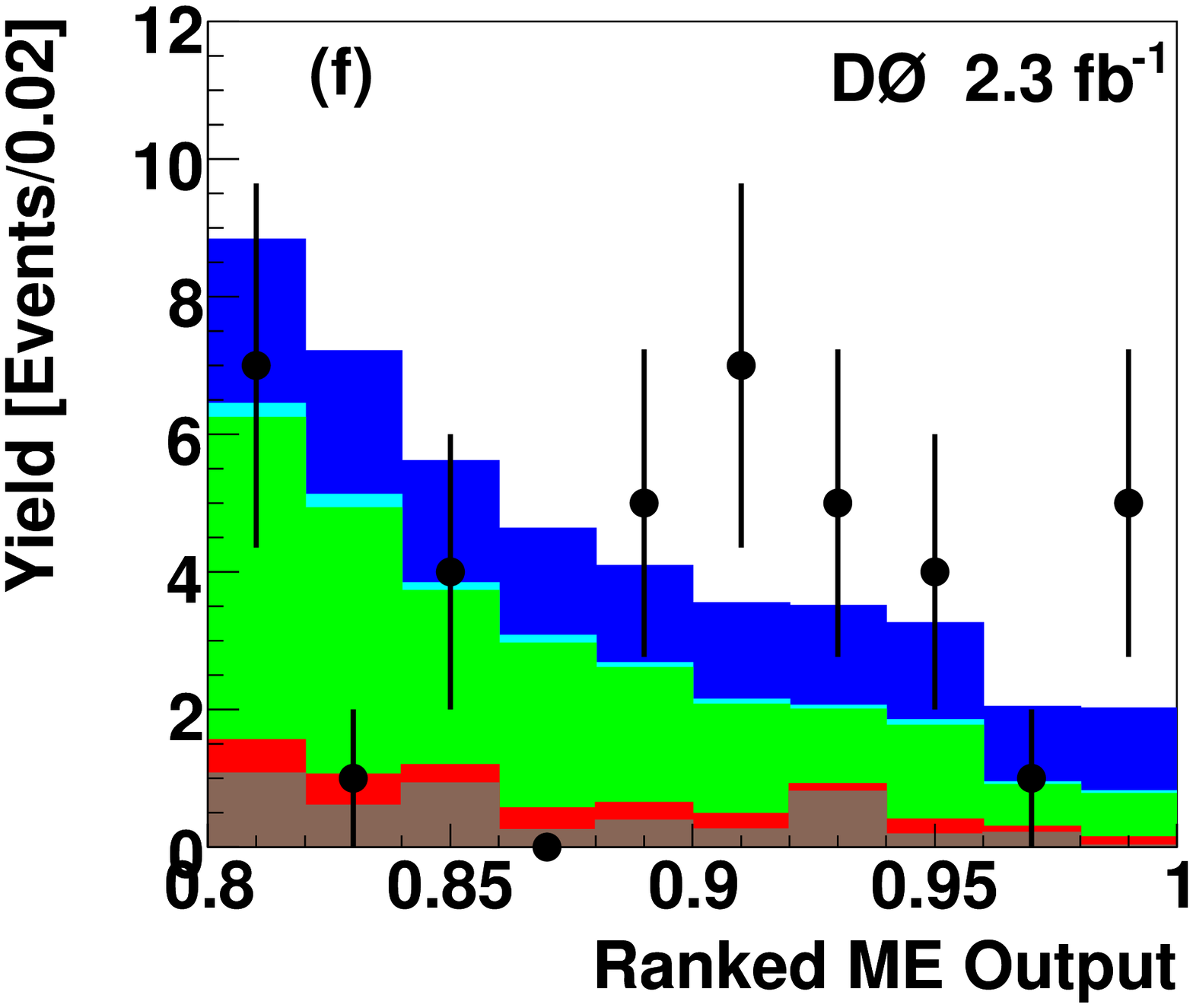}
\vspace{-0.3in}
\caption{Comparison of the signal and background models to data, for the BDT discriminant (a,b),
the BNN discriminant (c,d), and the ME discriminant (e,f), for the full discriminant range 
(a,c,e) and the signal region (b,d,f).
The bins have been ordered by their expected $t$-channel signal:background ratio 
and $t$-channel and $s$-channel single top distributions are normalized to the 
measured cross sections.}
\label{fig:BDTBNNME}
\end{figure}

%
We apply three independent multivariate analysis techniques to separate the
small $t$-channel single top signal from the large backgrounds,
based on boosted decision trees
(BDT)~\cite{decision-trees,bdt-benitez,bdt-gillberg}, Bayesian neural
networks (BNN)~\cite{bayesianNNs,bnn-tanasijczuk}, and the matrix
element (ME) method~\cite{matrix-elements,me-pangilinan}. These techniques
and their application are described in detail in Ref.~\cite{d0-prd-2008}.
For this analysis we use the same set of variables as in the observation
Letter~\cite{singletop-obs-d0}.
However, only $t$-channel single top events are considered signal during the
optimization, whereas $s$-channel single top events are included in the background,
normalized to the SM expectation. Fig.~\ref{fig:BDTBNNME} shows comparisons
between the $t$-channel signal, the background model, and data for the three 
individual discriminants.
~

\begin{figure}[!h!tbp]
\centering
\vspace{-0.1in}
\hspace{-0.1in}
\includegraphics[width=0.245\textwidth]{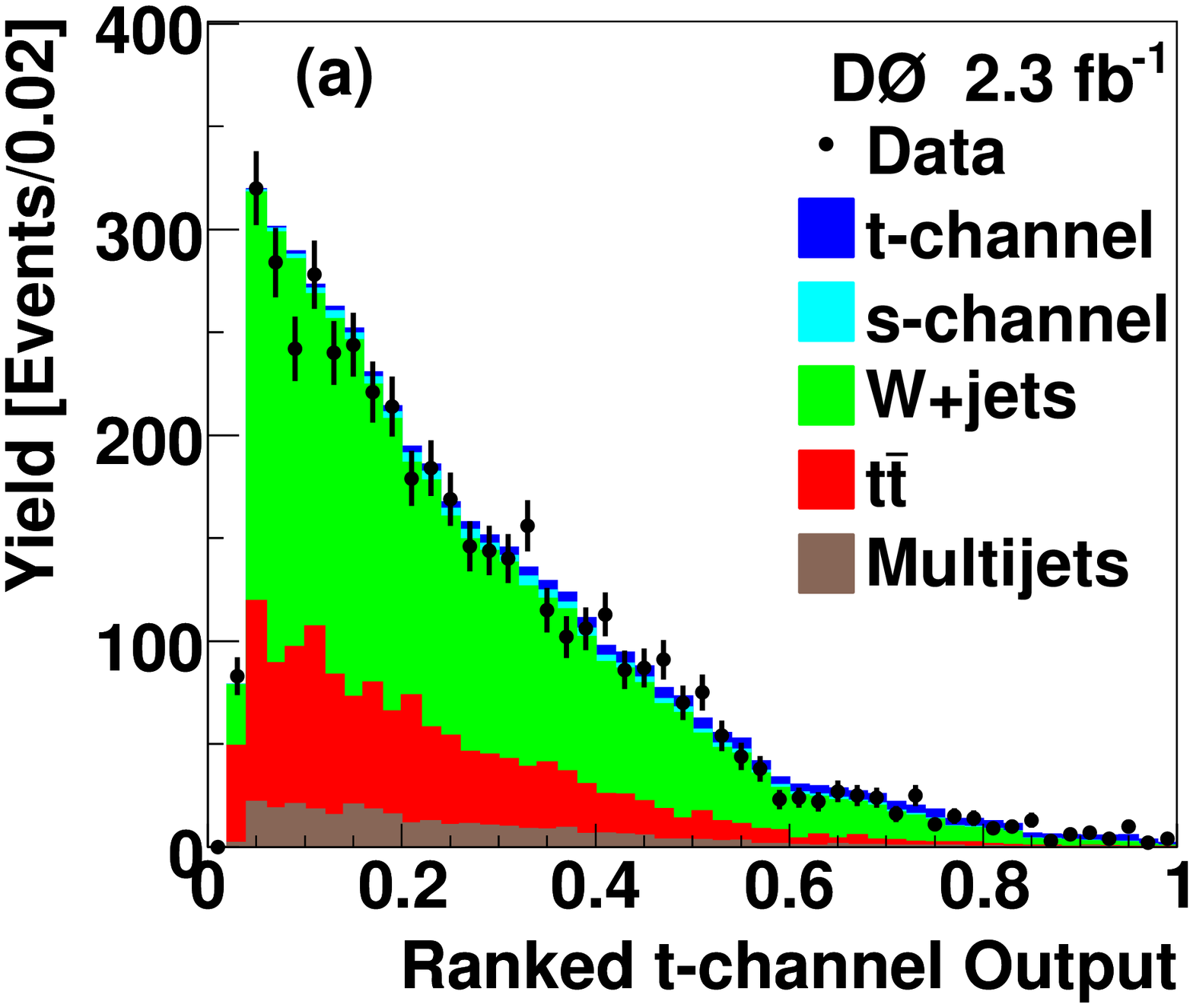}
\hspace{-0.1in}
\includegraphics[width=0.245\textwidth]{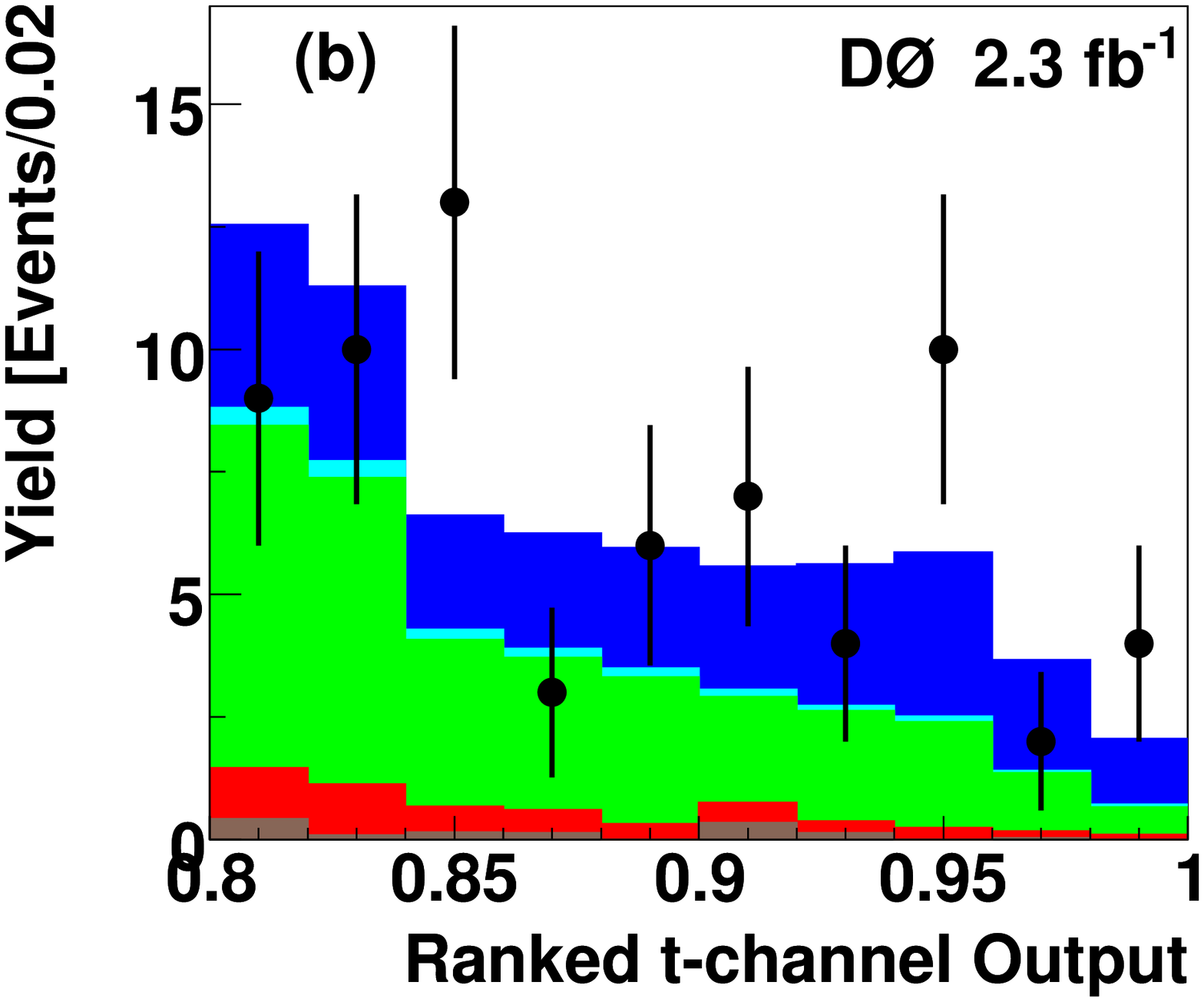}
\vspace{-0.3in}
\caption{Comparison of the signal and background models to data for the combination discriminant
output, for the full range (a) and only the signal region (b). The bins have been ordered by
their expected $t$-channel signal:background ratio and $t$-channel and $s$-channel single top 
distributions are normalized to the measured cross sections.}
\label{fig:discriminant}
\end{figure}

%
The three multivariate techniques use the same data sample but are not
completely correlated. Their combination leads to increased
sensitivity and a more precise measurement of the cross section. We
achieve this by training a combination BNN which uses the three individual discriminant 
outputs as inputs. Fig.~\ref{fig:discriminant} shows the combination
discriminant output for data superimposed on the background and signal
models. 
~

\begin{figure}[!h!tbp]
\centering
\hspace{-0.1in}
\includegraphics[width=0.245\textwidth]{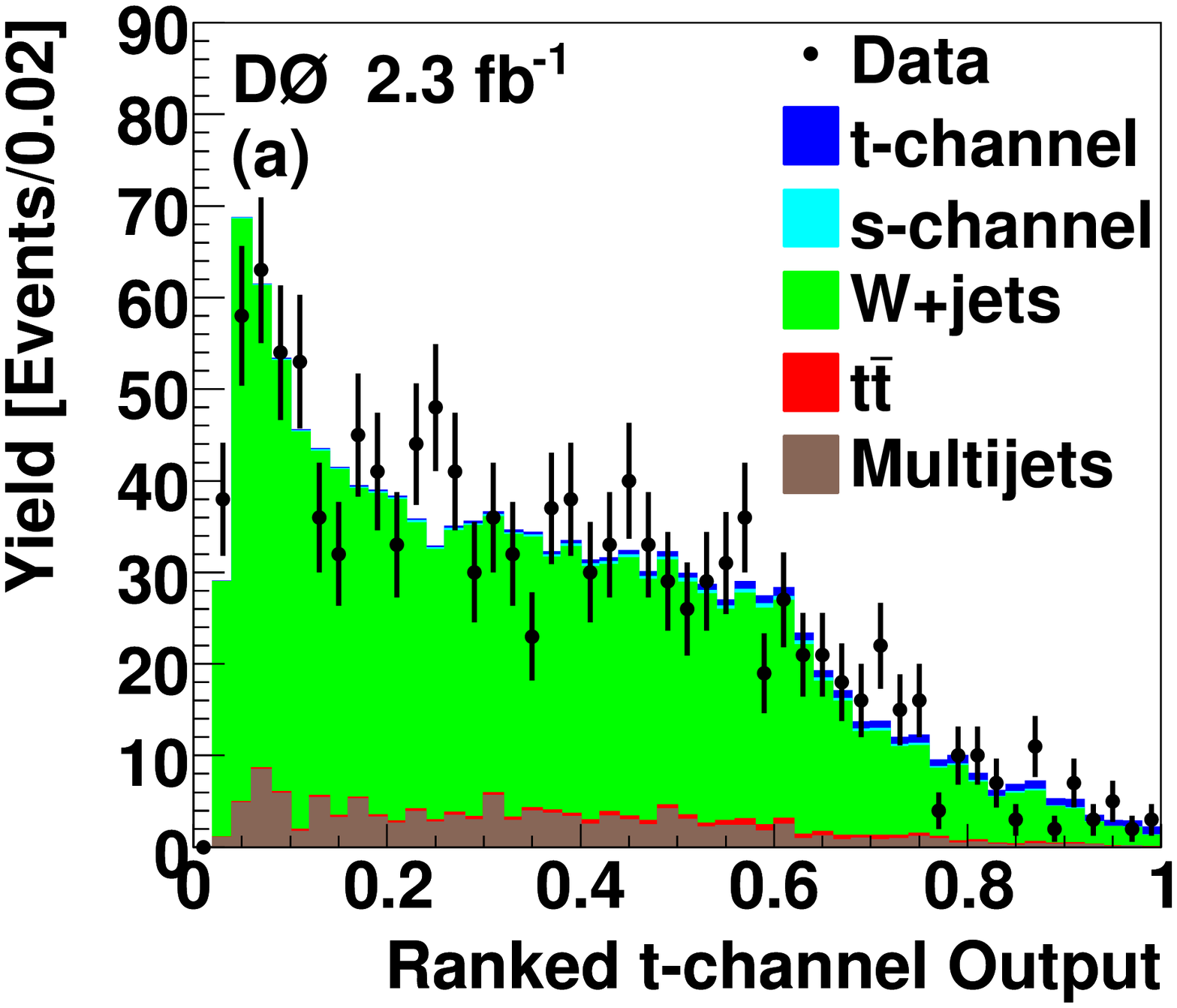}
\hspace{-0.1in}
\includegraphics[width=0.245\textwidth]{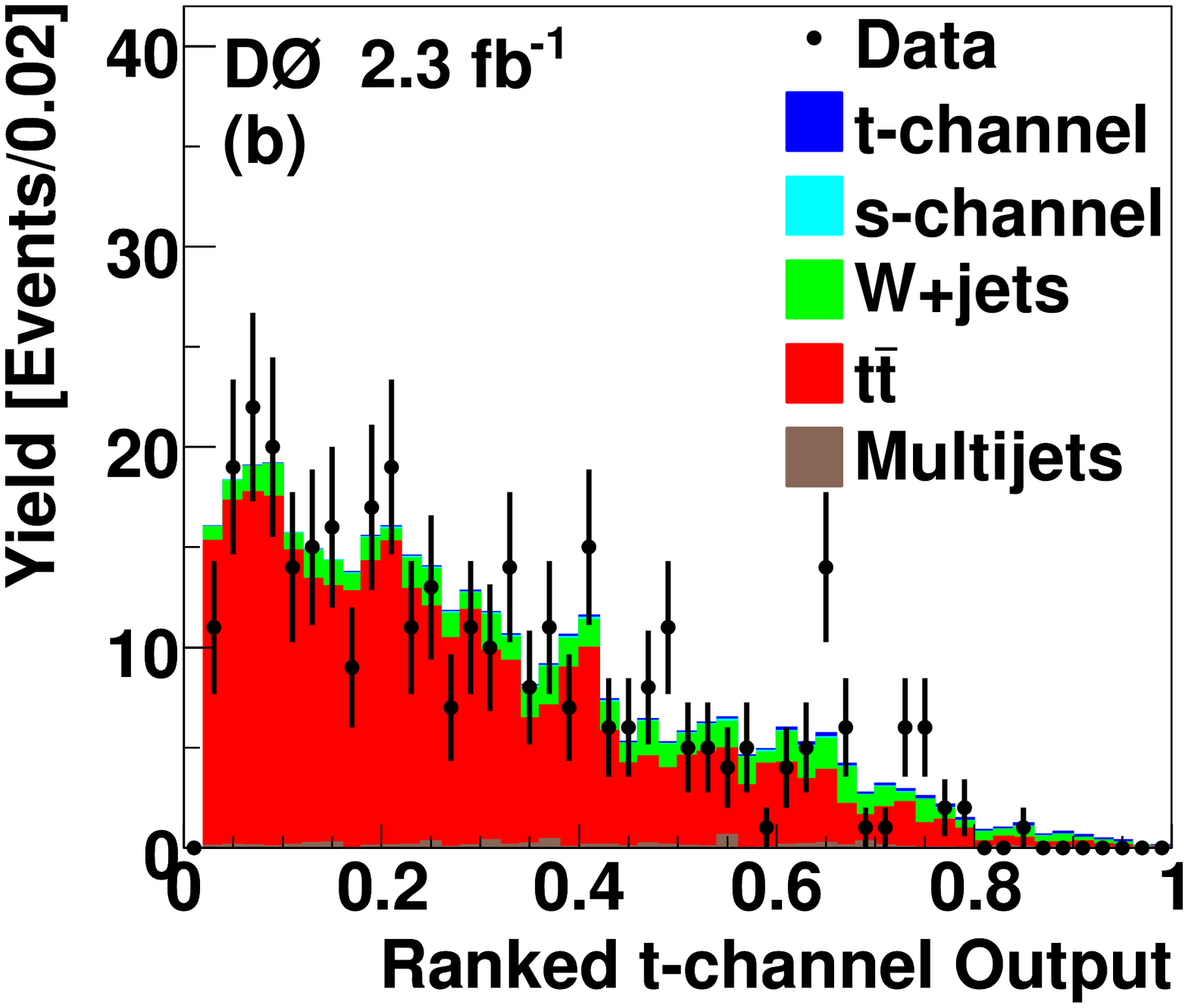}
\vspace{-0.3in}
\caption{Comparison of the background model to data for the ranked combination
output, for a $W$+jets (a) and a $\ttbar$ (b) dominated control sample.}
\label{fig:xcheck}
\end{figure}

We verify the accurate modeling of the data in background-dominated control regions
for the two main background categories. Fig.~\ref{fig:xcheck}a shows the $t$-channel 
discriminant in a $W$+jets dominated sample of 2-jet, 1-tag events with $H_T<175$~GeV. 
Fig.~\ref{fig:xcheck}b shows the $t$-channel discriminant
in a $\ttbar$ dominated sample of 4-jet, 1-tag or 2-tag events with $H_T>300$~GeV.
~  
These studies confirm that backgrounds are
well-modeled across the full range of the discriminant output.

\begin{figure}[!h!btp]
\centering
\includegraphics[width=0.45\textwidth]{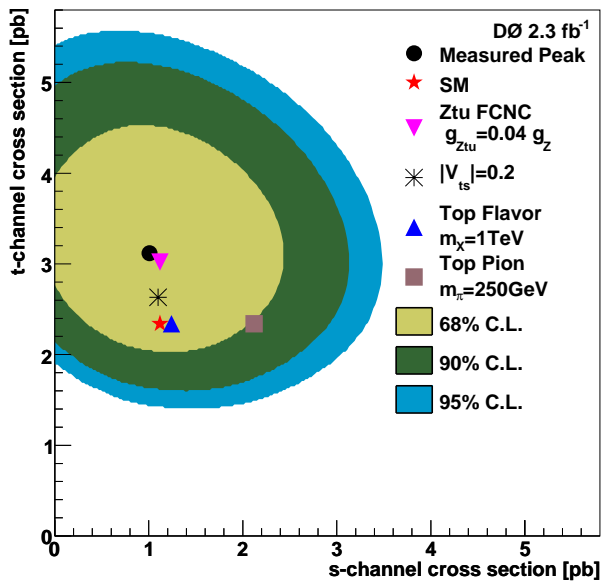} 
\vspace{-0.1in}
\caption{Posterior probability density for $t$-channel and $s$-channel single top quark 
production in contours of equal probability density. Also shown are the measured cross section, 
SM expectation, and several representative new physics scenarios~\cite{Tait:2000sh,Alwall:2007}.
}
\label{fig:posterior_2D}
\end{figure}

We use a Bayesian statistical analysis~\cite{bayes-limits} to measure the
production cross sections. In a first step we compute the two-dimensional
posterior probability density as a function of both $t$-channel and $s$-channel single top
quark cross sections. The combination  discriminants for 
$t$-channel and $s$-channel single top, remaining background, and data are used to build a 
binned likelihood as a product over all analysis channels and bins.  We assume a
Poisson distribution for the observed counts, and flat prior probabilities for positive
values of the $t$-channel and $s$-channel signal cross sections. Systematic uncertainties 
are described by Gaussian priors, and their correlations amongst all bins in all channels are
preserved. The posterior probability density is 
shown in Fig.~\ref{fig:posterior_2D}. 
Also shown are the SM expectation as well as several representative new physics models to
illustrate the sensitivity of this analysis. Dedicated searches should be able to address
flavor-changing neutral currents with a $Z$~boson coupling to the top and up 
quark with a strength of 4\% of the SM coupling~\cite{Tait:2000sh} or a top-color model with a 
$t\bar{b}$ bound state (Top Pion)  with a mass of $m_\pi=250$~GeV~\cite{Tait:2000sh},
while a 4-quark-generations scenario with CKM matrix element $|V_{ts}|=0.2$~\cite{Alwall:2007} 
or a top-flavor model with new heavy bosons at a scale $m_x=1$~TeV~\cite{Tait:2000sh} will be
more challenging to identify and might have to wait for LHC studies.

In a second step we obtain the $t$-channel posterior probability density from the 
two-dimensional posterior in Fig.~\ref{fig:posterior_2D} by integrating over the $s$-channel 
axis, thus not making any assumptions about the value of the $s$-channel cross section. We have 
analyzed ensembles of pseudo-datasets generated at several different $t$-channel and 
$s$-channel cross sections to verify the linearity of the measured $t$-channel cross section 
and its independence of the input $s$-channel cross section.
From the $t$-channel posterior we extract the cross section and uncertainty for $t$-channel 
single top quark production as $3.14^{+0.94}_{-0.80}$~pb. We similarly extract the $s$-channel
cross section as $1.05 \pm 0.81$~pb by integrating over the $t$-channel axis.

We compute the significance of the $t$-channel cross section measurement using pseudo-datasets 
generated from the background model (including SM $s$-channel single top) and taking all 
systematic uncertainties into account in a log-likelihood-ratio 
approach~\cite{Aaltonen:2009jj,Amsler:2008zzb}. 
For each pseudo-dataset we calculate the ratio of the probabilities for two hypotheses: 
that the pseudo-dataset is described by the background model only (including SM $s$-channel), 
and that it is described by SM 
$t$-channel single top plus backgrounds. We measure the p-value by counting the 
fraction of background-only pseudo-datasets with a ratio that is more signal-like 
than the one observed in data. 
The observed p-value is  $8.0\times 10^{-7}$,
corresponding to a Gaussian significance of 4.8$\sigma$, and the expected p-value is
$9.7\times 10^{-5}$, corresponding to a Gaussian significance of 3.7$\sigma$.
\begin{figure}[!h!btp]
\centering
\vspace{-0.1in}
\hspace{-0.1in}
\includegraphics[width=0.245\textwidth]{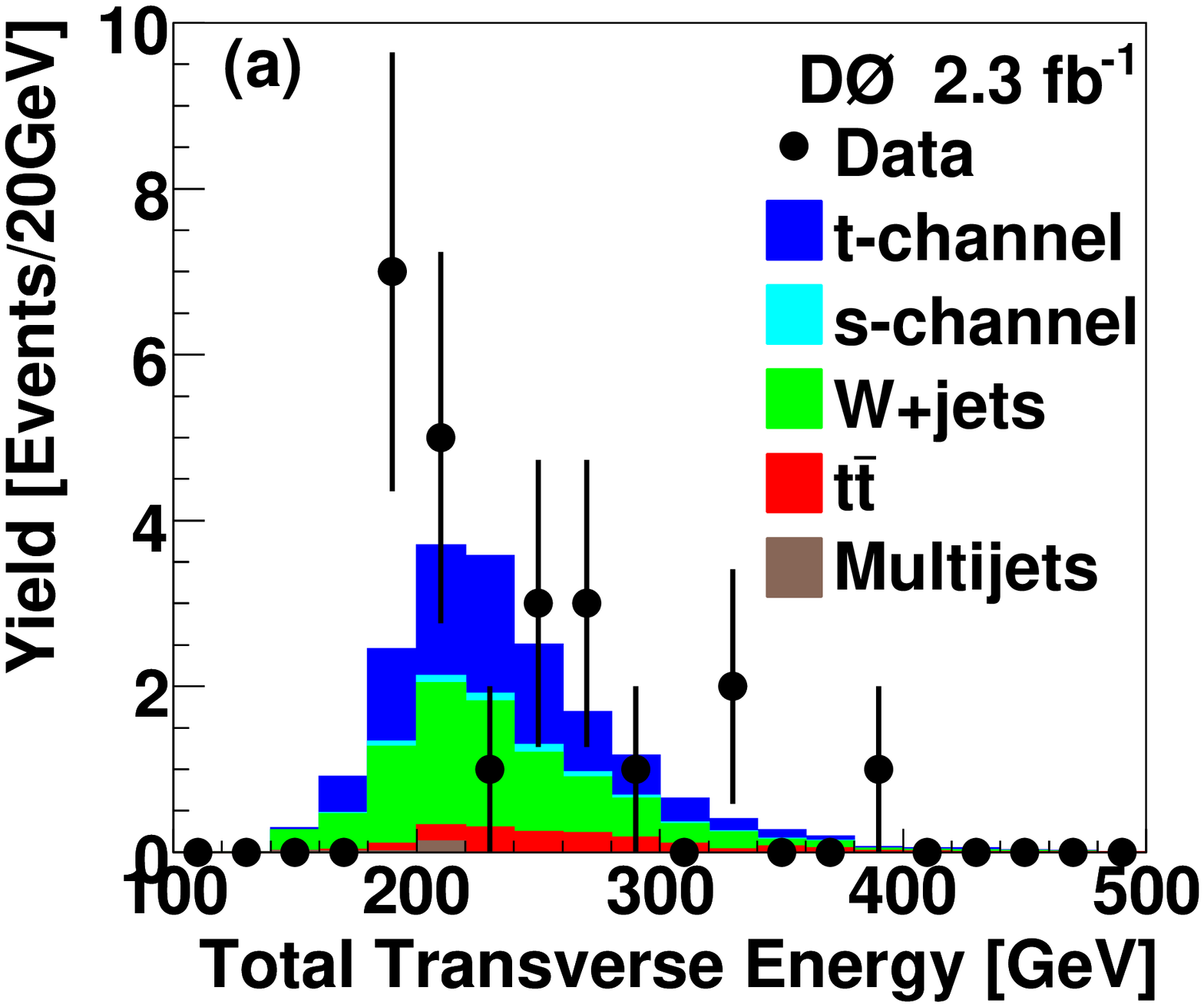} 
\hspace{-0.1in}
\includegraphics[width=0.245\textwidth]{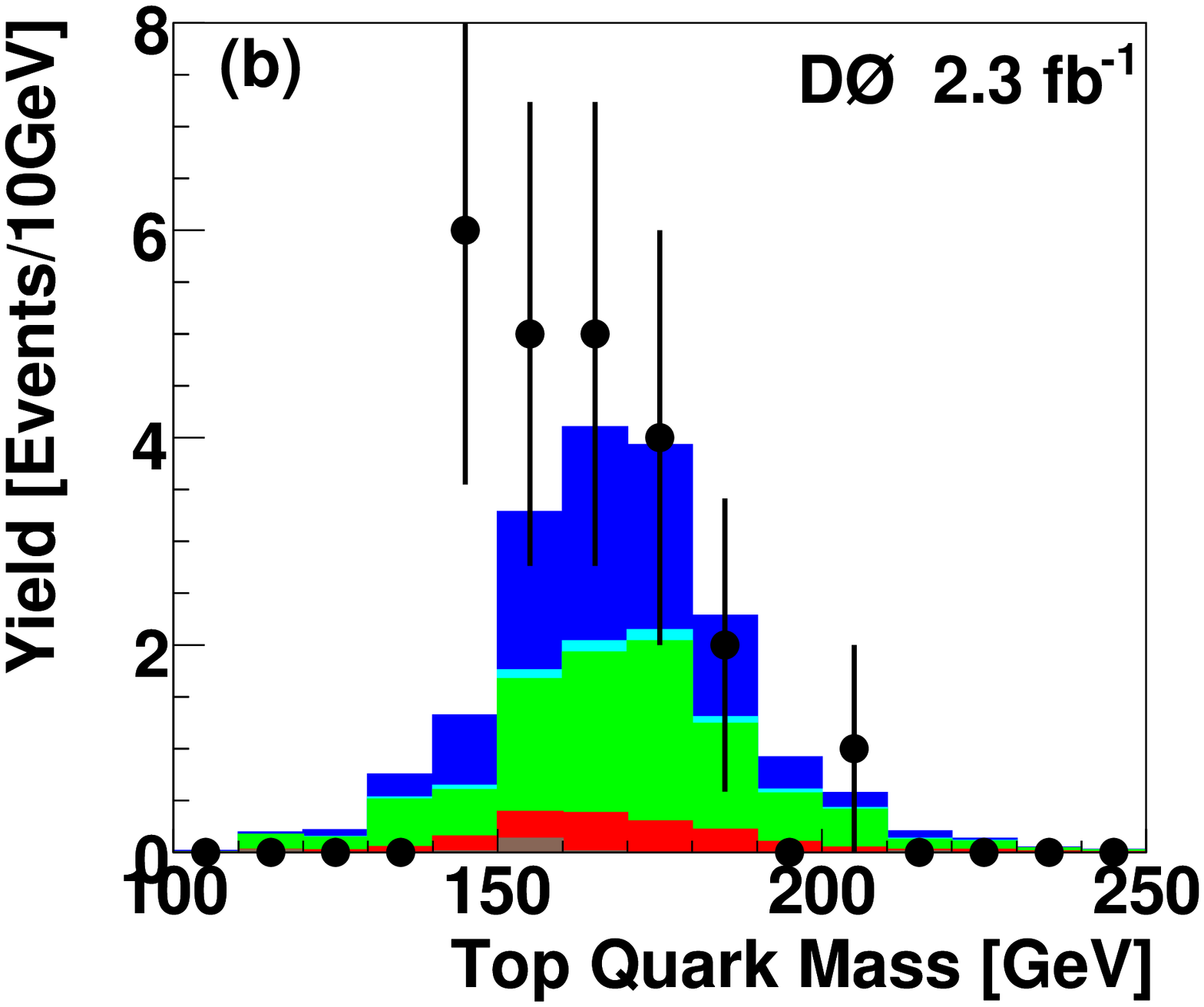} 
\\ \hspace{-0.1in}
\includegraphics[width=0.245\textwidth]{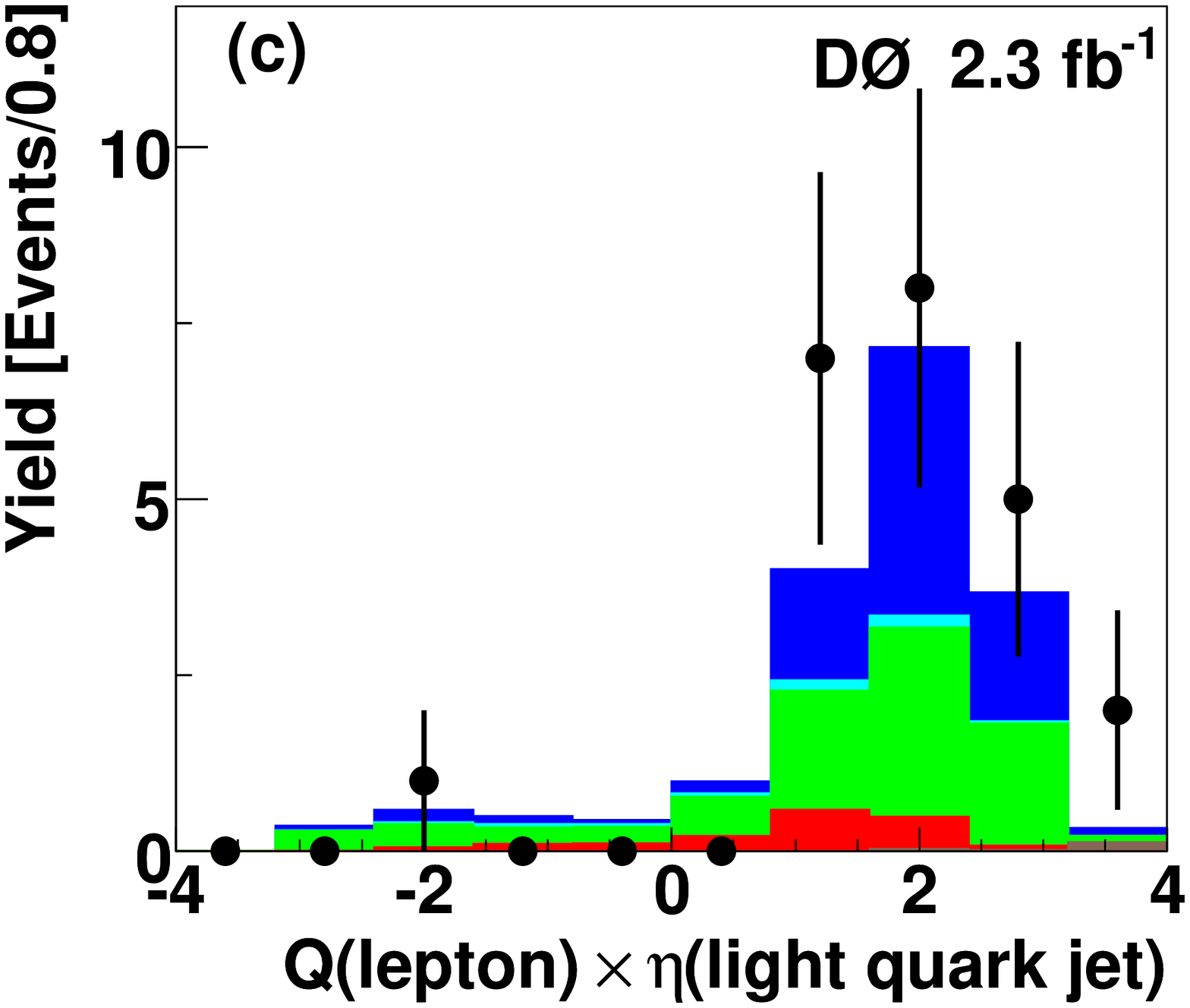} 
\hspace{-0.1in}
\includegraphics[width=0.245\textwidth]{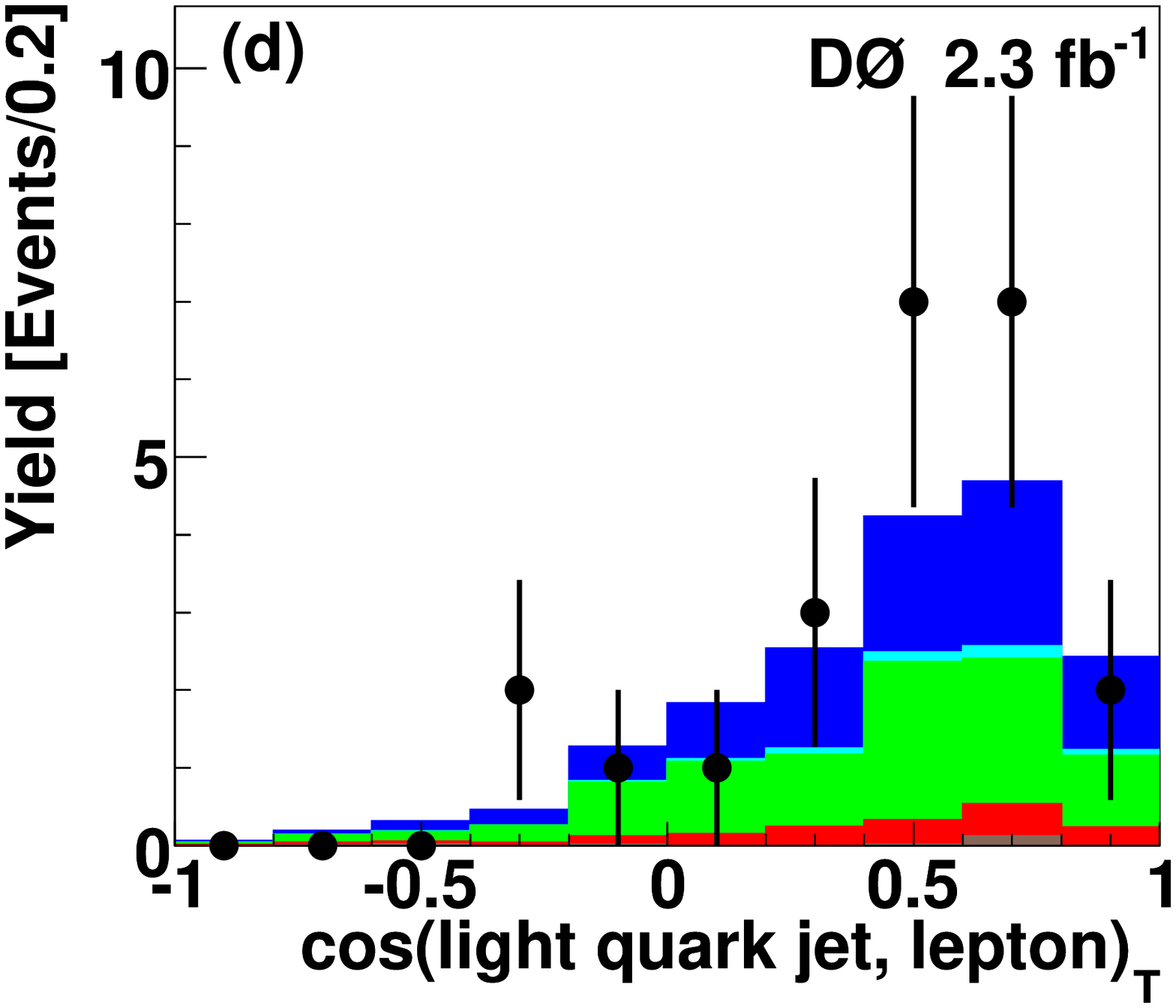} 
\vspace{-0.2in}
\caption{$H_T$ (a), 
reconstructed top quark mass (b), light quark
jet pseudorapidity multiplied by lepton charge (c), and $t$-channel
top quark spin correlation (d, see text) for events with a ranked combination
output $>0.91$. The $t$-channel and $s$-channel contributions have been
normalized to their measured cross sections.}
\label{fig:distributions}
\end{figure}

We have checked the consistency of the observed signal with SM $t$-channel events
in several kinematic distributions. Fig.~\ref{fig:distributions} shows comparisons
between the observed data, the background model, and the $t$-channel signal for four
different kinematic distributions for events with a ranked discriminant output $>0.91$.
Shown are four important kinematic variables for $t$-channel single top quark production:
$H_T$; the reconstructed top quark mass; 
the lepton charge multiplied by the pseudorapidity of the leading non-$b$-tagged jet (cf.
Fig.~\ref{fig:feynman}b); 
and the $t$-channel spin correlation in the optimal basis~\cite{Mahlon:1996pn,Cao:2005pq}, 
i.e. the cosine of the angle between the light quark jet and the lepton, both
in the reconstructed top quark rest frame. While the background shapes resemble the signal
in the high ranked discriminant output region, the presence of the $t$-channel signal is 
nevertheless clearly evident in each distribution.

In summary, we have presented the first direct evidence of the $t$-channel mode of 
single top quark production using 2.3~fb$^{-1}$ of data at the D0 experiment. 
We measure a $t$-channel cross section of $3.14^{+0.94}_{-0.80}$~pb and a $s$-channel
cross section of $1.05 \pm 0.81$~pb. The measured cross sections are consistent with the SM 
expected values. The observed $t$-channel signal corresponds to an
excess over the predicted background with a significance of $4.8\,\sigma$.

%
We thank the staffs at Fermilab and collaborating institutions, 
and acknowledge support from the 
DOE and NSF (USA);
CEA and CNRS/IN2P3 (France);
FASI, Rosatom and RFBR (Russia);
CNPq, FAPERJ, FAPESP and FUNDUNESP (Brazil);
DAE and DST (India);
Colciencias (Colombia);
CONACyT (Mexico);
KRF and KOSEF (Korea);
CONICET and UBACyT (Argentina);
FOM (The Netherlands);
STFC and the Royal Society (United Kingdom);
MSMT and GACR (Czech Republic);
CRC Program, CFI, NSERC and WestGrid Project (Canada);
BMBF and DFG (Germany);
SFI (Ireland);
The Swedish Research Council (Sweden);
CAS and CNSF (China);
and the
Alexander von Humboldt Foundation (Germany).
%
\vspace{-0.1in}


\end{document}